\documentclass[preprint,prd]{revtex4-1}
\usepackage{color}
\usepackage{graphicx}
\usepackage{epsfig}
\usepackage{subfig}
 \usepackage[font=small]{caption}

\usepackage{graphicx}
\usepackage{dcolumn}
\usepackage{bm}

\begin{document}

\title{ Gravity and Electromagnetism  with $Y(R) F^2$-type Coupling and Magnetic Monopole Solutions  }

\author{ \"{O}zcan SERT}
\email{osert@pau.edu.tr}
 \affiliation{Department of Physics,
Pamukkale
University, 20070,  K{\i}n{\i}kl{\i},  Denizli, T\"{u}rkiye}


\date{\today}

\begin{abstract}

 \noindent
 We   investigate    $ Y(R) F^2 $-type  coupling of electromagnetic fields   to  gravity.
  After we  derive field equations  by a first order variational principle
   from  the Lagrangian formulation of the non-minimally    coupled
theory, we look for   static,
spherically symmetric, magnetic  monopole solutions.
We point out that the solutions  can provide possible geometries which may  explain the flatness of  the observed rotation curves of galaxies.



\end{abstract}

\pacs{Valid PACS appear here}
\maketitle


\def\ba{\begin{eqnarray}}
\def\ea{\end{eqnarray}}
\def\w{\wedge}



\section{Introduction}

\noindent

We explore  the gravitational model which involves a non-minimal coupling between a
 function of curvature scalar and electromagnetic fields in $Y(R)F^2$ form.
The models with $RF^2$-type couplings were investigated in \cite{prasanna,horndeski,drummond,buchdahl,dereli1,muller-hoissen,balakin1,balakin2,balakin3,dereli2,Bamba5,Bamba6,Liu}   in order to obtain more information about   the modified gravity and electromagnetism.
Later, they were extended to the $R^mF^2$-type  couplings \cite{lambiase} to seed and grow the primordial magnetic fields in the universe.
On the other hand,    late-time acceleration mechanism of the universe has not been  well-established in modified $f(R)$ gravity (for review see \cite{capozziello}).
Also,  in the presence of the electromagnetic fields, the modified  $f(R)$ gravity  has  some black hole solutions  which  are not asymptotically flat  \cite{Mazharimousavi}.
Hence the natural extensions of $ f(R)$ gravity to the models, $f(R)$-Maxwell  \cite{bamba1,bamba2},  $f(G)$-Maxwell \cite{setare}, $f(R)$-Yang-Mills  \cite{bamba3} and $f(G)$-Yang-Mills gravity
 \cite{banijamali}
 were studied in order to explain the late-time acceleration and  inflation of  the universe.  Moreover, there are works remarking that
the rotational curves of test particles gravitating around galaxies may be realized by considering   a general non-minimal $f(R)$-matter couplings \cite{harko1,harko2,nojiri}.

Even though other modified $f(R)$ theories, involving a curvature scalar, exist,
non-minimal couplings with electromagnetic field need further investigations, in
order to show new relationships among gravity, electromagnetism and/or
electromagnetic duality.
 Furthermore, it is important to observe that the non-minimal couplings arise from one-loop vacuum   polarization effects in  quantum electrodynamics
of  the photon effective action in  a curved space-time \cite{drummond}.
Then, the non-minimal couplings that break the conformal invariance of electromagnetic field give rise to electromagnetic quantum fluctuations at the inflationary stage,
which lead  the inflation \cite{turner,mazzitelli,lambiase2,raya,campanelli}.  The  scale of the fluctuations can be stretched towards outside the Hubble horizon because of the inflation at that time
and they lead to classical fluctuations.
Therefore, the non-minimal couplings can be the reason of inflation and the large scale magnetic fields observed in clusters of galaxies\cite{kim1,kim2,clarke,turner2,bamba1}.
Also, there are some exotic scenarios  for producing the galactic magnetic field from the magnetic monopoles \cite{turner2}.
   Hence, we emphasize that  it is  important to find spherically symmetric  magnetic solutions consistent with observations from solar system to cosmological scales.


 In the present paper, we proceed to investigate the
non-minimal couplings of gravitational and electromagnetic fields
obtaining the magnetic solutions. We first discuss   a
non-minimally coupled Einstein-Maxwell theory. Then  we derive the
gravitational  field equations by a first order variational
principle using the method of Lagrange multipliers and algebra of
exterior differential forms, which is  in a way independent of the
choice of local coordinates. Consequently, we find  solutions
recovering   the metric functions in \cite{dereli3,dereli4} in the
presence of magnetic field.
  Thus,
  we point out that the solutions  are asymptotically flat  for some  values of the parameters in the model and
  they  can provide possible geometries which may  explain the flatness of  the observed rotational curves of galaxies.

\section{Field Equations of the Non-minimally Coupled Theory} \label{model}

\bigskip

\noindent We will derive the field equations of the non-minimal theory by a variational
principle from an action
\begin{equation}
        I[e^a,{\omega^a}_b,F] = \int_M{L} = \int_M{\mathcal{L}^*1},
        \nonumber
\end{equation}
where  $\{e^a\}$ and ${\{\omega^a}_b\}$ are the fundamental
gravitational field variables and   $F=dA$ is the electromagnetic
field 2-form.  The space-time metric $g = \eta_{ab} e^a \otimes
e^b$ has the signature $(-+++)$ and we fix the orientation by setting
$*1 = e^0 \w e^1 \w  e^2 \w e^3 $.  Torsion 2-forms $T^a$ and
curvature 2-forms $R^{a}_{\; \; b}$ of spacetime are given in
the Cartan-Maurer structure equations
\begin{equation}
T^a = de^a + \omega^{a}_{\;\;b} \w e^b , \nonumber
\end{equation}
\begin{equation}
R^{a}_{\;\;b} = d\omega^{a}_{\;\;b} + \omega^{a}_{\;\;c} \w \omega^{c}_{\;\;b} . \nonumber
\end{equation}
We consider  the following  Lagrangian density 4-form;
 \ba \label{action}
  L =  \frac{1}{2\kappa^2} R*1 -\frac{1}{2}Y(R) F\w *F +  T^a \w \lambda_a ,
   \label{Lagrange}
   \ea
where
 $\kappa^2 = 8\pi G$ is  Newton's universal gravitational constant $(c=1)$ and $R$ is the curvature scalar which can be found by applying interior product $\iota_a $ twice to the curvature tensor $R_{ab}$ 2-form.  This  Lagrangian density  involves Lagrange multiplier 2-form $\lambda_a$  whose variation imposes the zero-torsion constraint  $T^a=0$.

We use the shorthand notation $ e^a \wedge e^b \wedge \cdots =
e^{ab\cdots}$, and  $\iota_aF =F_a, \  \  \iota_{ba} F =F_{ab}, $ \   $ \iota_a {R^a}_b =R_b, \  \   \iota_{ba} R^{ab}= R $.
   The field equations are obtained by considering the independent variations of
   the action with respect to  $\{e^a\}$,
   ${\{\omega^a}_b\}$ and $\{F\}$.  The electromagnetic field components are read  from the expansion $F = \frac{1}{2} F_{ab} e^a  \w  e^b$.
We will be working with the unique metric-compatible
Levi-Civita connection.
\medskip

\noindent The infinitesimal variations of the total Lagrangian
density $L $ (modulo a closed form) are given by
\begin{eqnarray}\label{generaleinsteinfe1}
&& \dot{L}  = \frac{1}{2 \kappa^2} \dot{e}^a \w R^{bc}
\w *e_{abc} +  \dot{e}^a \w \frac{1}{2} Y(R)   (\iota_a F \w *F - F \w \iota_a *F)   +  \dot{e}^a \w D \lambda_a \nonumber
\\ & &
 + \dot{e}^a \w     Y_R     (\iota_a R^b)\imath_b( F \w  *F  )   + \frac{1}{2} \dot{\omega}_{ab} \w  ( e^b
\w \lambda^a - e^a \w \lambda^b)
  \nonumber \\
& & + \dot{\omega}_{ab} \w  {\Sigma}^{ab}
 -\dot{ F} \w Y(R)  *F   +
\dot{\lambda}_a \w T^a .
\end{eqnarray}
where  $Y_R = \frac{dY}{dR}$, and the angular momentum tensor
\begin{eqnarray}\label{sigmaab1}
 {\Sigma}^{ab} &=&  \frac{ 1}{2} D   \imath^{ab}[ Y_R F  \w *F ].
   \end{eqnarray}
The Lagrange multiplier 2-forms $\lambda_a$ are solved  uniquely
from the connection variation equations
 \begin{eqnarray}\label{lambdaaeb}
 e_a\w \lambda_b -  e_b \w \lambda_a = 2{\Sigma}_{ab},
 \end{eqnarray}
by applying the  interior product operator twice as
\begin{eqnarray}\label{lambdaaeb2}
\lambda^a &=&  2\imath_b   {\Sigma}^{ba}  +\frac{1}{2}
\imath_{bc}  {\Sigma}^{cb}\wedge e^a.
\end{eqnarray}
We substitute the $ \lambda_a$\rq{}s into the $\dot{e^a}$ equations and after some simplifications we find the  Einstein field equations for the extended theory as
\begin{eqnarray}\label{einstein}
&&   \frac{1}{2 \kappa^2}  R^{bc}
\w *e_{abc} +  \frac{1}{2} Y  (\iota_a F \w *F - F \w \iota_a *F)   + Y_R  (\iota_a R^b)\iota_b( F \w *F )
 \nonumber
\\
&& + \frac{1}{2}  D [ \iota^b D(Y_R F_{mn} F^{mn} )]\wedge *e_{ab}
 =0   ,
\end{eqnarray}
while the Maxwell equations are
\begin{equation}\label{maxwell1}
d(Y * F) = 0, \  \  \   \  dF=0.
\end{equation}
  We note that our action (\ref{action})  and the field equations (\ref{einstein})-(\ref{maxwell1})
   when written out explicitly in any local coordinate system are equivalent to  the action in \cite{bamba2}.
    However, our variational derivation involving Lagrange multipliers is given in a way independent of the choice of coordinates.


%
 \section{Static, Spherically Symmetric,  Magnetic Solutions }

\noindent We consider (1+3)-dimensional static, spherically symmetric  solutions to the non-minimal model which are given by the metric
\begin{equation}\label{metric}
              g = -f(r)^2dt^2  +  f(r)^{-2}dr^2 + r^2d\theta^2 +r^2\sin(\theta)^2 d \phi^2
\end{equation}
 and the following   electromagnetic tensor $ F$ which has only magnetic component;
 \begin{eqnarray}\label{electromagnetic1}
 F   &=& B(r) r^2  \sin(\theta) d\theta\w  d\phi=  B(r)e^2 \w e^3.
\end{eqnarray}

 The   non-minimally coupled field  equations  (\ref{einstein}) give us   the following system of equations  for the metric (\ref{metric}) together with the magnetic field 2-form (\ref{electromagnetic1}):
 \begin{eqnarray}\label{e1}
  && \frac{1}{\kappa^2}(\frac{{f^2}\rq{}}{r}  + \frac{f^2-1 }{r^2} ) +  Y_R B^2 (\frac{ {{f^2}\rq{}}\rq{} } {2} + \frac{{f^2}\rq{} }{r}  ) + \frac{1}{2} YB^2
  + [(B^2  Y_R )\rq{} f]\rq{}f  +  \frac{2}{r} f^2 (B^2 Y_R   )\rq{}=0,
  \nonumber \\
    &&
\frac{1}{\kappa^2}(\frac{{f^2}\rq{}}{r}  + \frac{f^2-1 }{r^2} ) +  Y_R B^2 (\frac{ {{f^2}\rq{}}\rq{} } {2} + \frac{{f^2}\rq{} }{r}  ) + \frac{1}{2} YB^2
 + (B^2  Y_R )\rq{} ( \frac{{f^2}\rq{}}{2} + \frac{2 f^2}{r} ) =0 ,\\
 &&
 \frac{1}{\kappa^2}(\frac{ {{f^2}\rq{} }\rq{} }{2}  + \frac{{f^2}\rq{}}{r} ) + Y_R B^2 ( \frac{  {{f^2}\rq{}} } {r} + \frac{{f^2-1} }{r^2}  ) - \frac{1}{2} YB^2
  +  [(B^2  Y_R )\rq{} f]\rq{}f  +  (B^2  Y_R )\rq{} ( \frac{{f^2}\rq{}}{2} + \frac{f^2}{r} )=0, \nonumber
 \end{eqnarray}
Here the curvature scalar is calculated as
\begin{eqnarray}
R=- {{f^2}\rq{}}\rq{} -\frac{4 }{r} {f^2}\rq{} -\frac{2}{r^2} ( f^2-1) .
\end{eqnarray}

    \subsection{Exact Solutions}

\noindent We note that the system of equations are reduced to a simpler form when we assume the condition
\begin{eqnarray}\label{dif3}
Y_R B^2=C
\end{eqnarray}
where $C$ is a constant to be fixed according to (\ref{dif5}).
Then,  the field equations (\ref{e1})  turn out  to be
\begin{eqnarray}\label{dif4}
\frac{1}{2}\left( {{f^2}\rq{}}\rq{} -\frac{2}{r^2}(f^2-1) \right) \left(  C -  \frac{1}{\kappa^2}   \right) + YB^2 &=&0 , \\
\frac{R}{2}( \frac{1}{\kappa^2} + C  )\label{dif5}
&=&0.
\end{eqnarray}

\noindent In this paper,  to recover  the previous  metric functions   obtained in \cite{dereli3,dereli4}   we deal with certain  non-minimal functions  $Y(R)$ that involve a new non-minimal coupling constant    $\lambda$ which can be related to cosmological constant.

\begin{itemize}
\item
First, we consider the inverse function of \cite{dereli3}, i.e. the following non-minimal coupling \cite{mazzitelli,Bamba7}:
\begin{eqnarray}\label{nonminimalsol1}
Y(R)=1-a_1\ln( \frac{R_\lambda}{R_0} )
\end{eqnarray}
where $R_\lambda(r)=R (r)+ 12\lambda$,   $ a_1  $ is a dimensionless  coupling
   constant,  $R_0$ is a constant with the same dimension as that of $R$.
 We note  that the non-minimal coupling constants $a_1$  and $ R_0 $  are related to  the strength  of  the  electromagnetic field-gravity couplings.

 As a phenomenological
approach to take these forms of $Y(R)$,
 in \cite{Elizalde}, it has been demonstrated  that a logarithmic
non-minimal gravitational coupling like this type appears in the
effective renormalization-group improved Lagrangian for an $SU(2)$
gauge theory in matter sector for a de Sitter background. For this
model, the time variation of the fine structure constant has
recently been examined in \cite{Bamba-Nojiri}.

     In this case we find the  following geometry and magnetic field:
\begin{eqnarray}\label{nonminimalsol2}
f^2(r) &=& 1-\frac{2M}{r}+\frac{a_1\kappa^2q^2}{r^2}\ln \frac{r}{r_0}  +\frac{\kappa^2q^2 ( 1+ 5a_1)} {4r^2} + \lambda r^2,   \hskip 1 cm for \ a_1 \neq 0,  \\
B(r) &=& \frac{q}{r^2},
\end{eqnarray}
where $r_0$ is an integration constant satisfying the relation $r_0^4=\frac{a_1\kappa^2 q^2}{R_0}$, $q$ and $M$ are respectively magnetic charge and mass of the gravitating object. Thus  the curvature scalar becomes
 \begin{eqnarray}
 R(r)&=& \frac{a_1 \kappa^2 q^2}{r^4} -12 \lambda.
\end{eqnarray}

\item Second, the inverse  of   \cite{dereli4}, i.e. the following non-minimal coupling \cite{drummond,mazzitelli,lambiase} which reads:
\end{itemize}
\begin{eqnarray}\label{nonminimalsol3}
Y(R)=1-( \frac{R_\lambda}{R_0} )^\beta
\end{eqnarray}

we find  the following metric function and magnetic field:
\begin{eqnarray}\label{nonminimalsol4}
f^2(r) &=& 1-\frac{2M}{r}+\frac{ \kappa^2q^2} {4r^2}  - \frac{a_2(\beta-1)^2}{4\beta(3\beta+1)}r^{\frac{2\beta+2}{\beta-1}} + \lambda r^2,   \hskip 1 cm for \ R_0 \neq 0, \  \  \beta \neq 0,1,-\frac{1}{3}  \\
B(r) &=& \frac{q}{r^2},
\end{eqnarray}
which give the curvature scalar
\begin{eqnarray}
 R(r)&=& a_2 r^{\frac{4}{\beta-1}} -12\lambda
\end{eqnarray}
where $a_2 =(\frac{{R_0}^{\beta}}{\kappa^2 q^2 \beta }
)^{\frac{1}{\beta-1}}$.

\noindent We note that
this system of equations (15-17) has consistent solutions for only the magnetic  field
$ B=\frac{q}{r^2}$   which can be obtained from the magnetic monopole (Dirac monopole)  potential
 \begin{eqnarray}\label{electromagnetic}
 A  &=& q (1-\cos(\theta) )  d\phi
\end{eqnarray}
which is  determined by the Gauss  integral
   \begin{eqnarray}
    \frac{1}{4\pi} {\int_{S^2}{ F }} =  \frac{1}{4\pi} {\int_{S^2}{ B(r) r^2 \sin \theta  d\theta \wedge d\phi}}=q.
    \end{eqnarray}

The analysis of the  solutions such as  horizons and asymptotic
behaviors can be found in \cite{dereli3,dereli4}. The metric
function   (23)   with $ \beta=-3 $   give us a Rindler
acceleration term which is  also obtained from a Dilaton-gravity
model in \cite{grumiller, grumiller2}  to explain some anomalies
such as the rotation curves of spiral galaxies and Pioneer anomaly
\cite{bertone,anderson}. Furthermore, the case with $ 0 <   -
{\frac{2\beta+2}{\beta-1}}  < 1$ has the term which is more
efficient in the large  scale region.  This case corresponds to an
attractive force  for $ \frac{a_2(\beta-1)^2}{4\beta(3\beta+1)} >0
$. Thus,   these solutions may explain the dark matter effects
like the formation of the galaxies in these ranges $-\frac{1}{3}<
\beta <0$ and $a_2 < 0$ or  $0< \beta <1$ and $a_2 > 0$.
 If dark matter is not an exotic matter,  the non-minimal couplings  give rise to such effects \cite{nojiri4}
  for   the  values of  the parameters in the above intervals depending on the physical system.

The metric (\ref{nonminimalsol4}) leads to the effective
potential;
\begin{eqnarray}
V_{eff} = -\frac{M}{r} + \frac{\kappa^2q^2}{8 r^2} -
\frac{a_2(\beta-1)^2}{8\beta(3\beta +1)} r^{\frac{2 \beta +2}{
\beta -1}}
\end{eqnarray}
for vanishing angular momentum and cosmological constant.
 This effective potential can be rewritten in the following form
\begin{eqnarray}
V_{eff} = -\frac{M}{r}  - \frac{r^{\tilde{\beta} -1
}}{{r_c}^{\tilde{\beta}}} + \frac{\kappa^2q^2}{8 r^2}
\end{eqnarray}
where $\tilde{\beta} = \frac{3\beta+1}{\beta-1} $. We notice that
for large scale and $\tilde{\beta} > 0 $ the last term vanishes
and our potential   reduces to the effective potential in
\cite{Capozziello2, Capozziello3}. More comprehensive analysis of
this case can be found  there.

In order to give more indications on how confront our results with
observational data we calculate the speed of rotational curves
 for the metric (\ref{nonminimalsol4}) from $v^2= r \frac{dV_{eff}}{dr}$

\begin{eqnarray}
v=\sqrt{ \frac{M}{r}- \frac{\kappa^2q^2}{4r^2} - \frac{a_2(\beta^2
-1)}{4 \beta(3\beta +1 )} r^{\frac{2\beta +2}{\beta -1}} }
\end{eqnarray}

\noindent  The graphs of  speed vs. radius are plotted  in FIG.
1-(a) for small galaxies ($10^8$ solar masses $\simeq 10^{46}$
Planck mass) and in FIG. 1-(b) for large spiral galaxies
($10^{11}$ solar masses $\simeq  10^{49}$ Planck mass), assuming
that  mass density is constant until $r=10^{54}$ and $r=2\times
10^{55}$, respectively,  then goes to zero. The graphs resemble to
the observational velocity profiles  see e.g. \cite{Capozziello4},
where $10^{-3} = $  300 km/s. Thus,  the large scale magnetic
field which is generated due to the breaking of the conformal
invariance of the electromagnetic field can be reasons of the  the
flatness of the rotational curves of galaxies for some parameter
values.

   \begin{figure}[h]{}
  \centering
    \subfloat[  ]{ \includegraphics[width=0.4\textwidth]{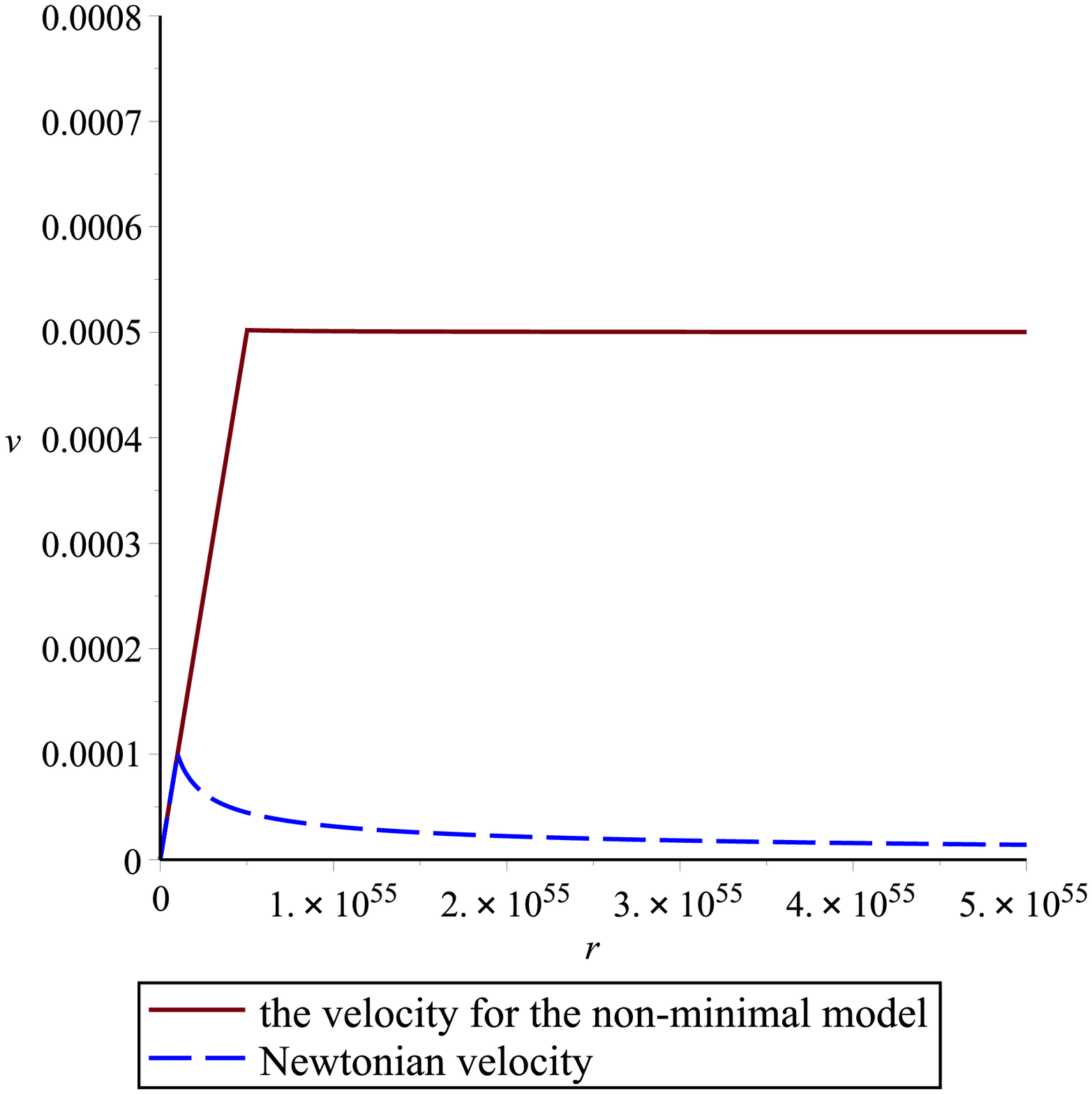} }
       \subfloat[ ]{ \includegraphics[width=0.4\textwidth]{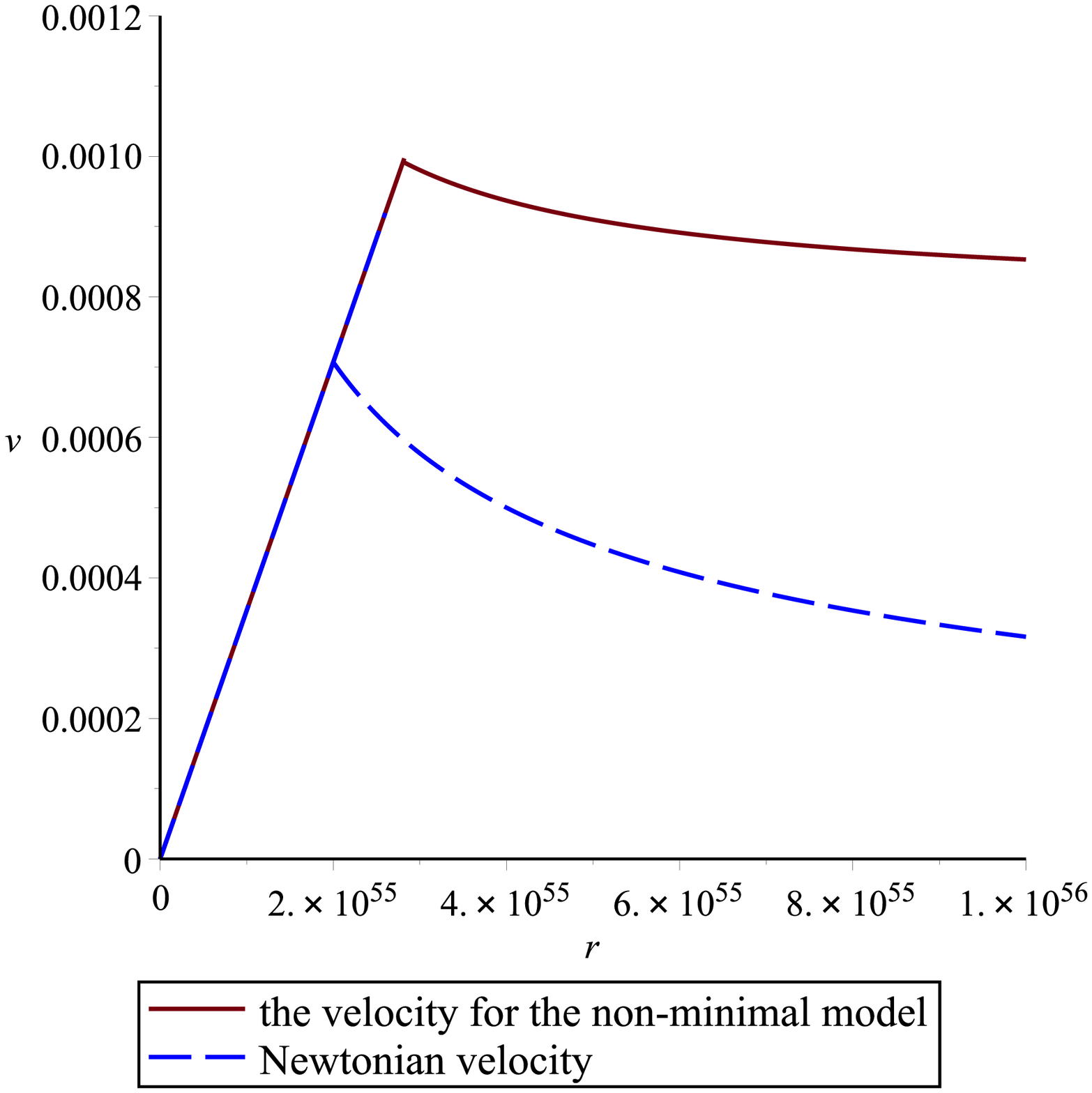}}  \\
  \parbox{6in}{\caption{{{{ \small The velocity profiles of rotational curves  for a) dwarf
  galaxy with $\beta=-1.000001=a_2=R_0$ and   $q^2\kappa^2=1$, b) large spiral galaxy  with $\beta=-1.000002512=a_2=R_0$ and   $q^2\kappa^2=1$.
                           }}}}}
          \end{figure}

 \noindent
It is relevant to point out that equations (\ref{nonminimalsol2})
and (\ref{nonminimalsol4}) show naked singularity solutions, for
some values of the involved parameters (see for example
\cite{dereli3}). Nevertheless, naked singularities, evaluated at
some future space-time points, could be revealed by an observer,
giving observational consequences
\cite{penrose,Luongo,Broda,Gasperini}. The naked singularities and
black holes could be observationally differentiated through their
gravitational lensing features (number of images, their
orientations, magnifications, and time delay,
etc.)\cite{virbhadra1,virbhadra2,virbhadra3,virbhadra4}.


\section{Conclusion}

\noindent We  have considered a   non-minimally $Y(R)F^2$-coupled
Einstein-Maxwell theory and looked for   static, spherically
symmetric and magnetic  solutions. After we find the reduced field
equations  for the theory, we choose the non-minimal $Y(R)$
function to recover the metric solutions which is obtained from
previous  studies with electric field \cite{dereli3,dereli4}. In
\cite{dereli3,dereli4}, those solutions are electrically charged
and the electric field changes sign due to polarization effects.
But, in this work, the magnetic monopole field is not modified  by
the non-minimal couplings. In addition, we obtain consistent
magnetic monopole solutions only for the $Y(R)$ function which is
equal to the inverse of that  in \cite{dereli3,dereli4}.

Meanwhile,  the pp-wave solutions of the non-minimal couplings in
$RF^2$ form can be found in \cite{dereli2}. Though it is difficult
to find an exact electric monopole solution to the model, an exact
magnetic monopole solution  was found for a special case in
\cite{balakin2}. In this case, Reissner-Nordstrom solution was
modified by $Q/r^4$ term. But it is difficult to find more general
modifications to the Reissner-Nordstrom-like solutions, such as
$r^\alpha$ or $ln(r)$. This work fills in this gap.

In the zero  magnetic charge limit,  the  model with the
non-minimal couplings  reduces to the Einstein gravity  which  has
some observational inconsistencies  such as   flatness of galactic
rotation curves and Pioneer anomaly.  But, in the case of
non-zero magnetic charge,  the  model leads to new solutions
(\ref{nonminimalsol2}) and (\ref{nonminimalsol4}).
   We can apply  these new solutions to  the behavior of  a test object  in   gravitational potential of a galaxy
    (like a star) or  the solar system (like a planet).
    In these cases,  the effective potential of the test object is  modified similarly as in  \cite{grumiller,grumiller2,nojiri4,Capozziello2,Capozziello3,Capozziello4}.
     Here the new  modifications  are generated from the total magnetic charge in this closed region.
 Future efforts can be spent to investigate if our model can show a unified dark matter behavior (see for instance \cite{gao,aviles}).
Moreover, the solutions which have  the  naked singularities  are
important  for their gravitational lensing features (number of
images, their orientations, magnifications, time delay, etc.) and
repulsive effects of gravity
\cite{penrose,virbhadra1,virbhadra2,virbhadra3,virbhadra4}.

  The non-minimal coupling of the curvature scalar with magnetic fields may not be
efficient in small scale, but it gives  important modifications
   to the gravity in astrophysical scale. Since the value of the coupling parameters depends on the system we describe, our metric does not necessarily spoil the solar system precision tests  and
    the non-minimal coupling
  can be   considered  as one of the  sources  for   shedding light on this cosmic  dark matter,  and vice versa.
On the other hand,    there is no observational evidence  on the existence of  magnetic monopoles in galaxies;
however,  it does not  exclude the possibility of their existence, and  it  may be  explained by  their scarcity.
Even if magnetic monopoles are rare, they may be very heavy.  Thus,  they may have important effects to the  dynamics of galaxies and make up the dark matter of galactic halos \cite{preskill}. But it is a challenging problem to detect these rare, heavy monopoles.
 Future experiments such as MOEDAL  \cite{moedal},  ANITA-II \cite{anita}, AMANDA-II \cite{amanda} and  IceCube \cite{icecube}
may  solve the  fundamental mystery of the universe.

\vskip 1cm

\section{Acknowledgement}

\noindent The author would like to thank  Muzaffer Adak and the
organizers of the $11^{th}$ Workshop on  {\it  Quantization
Dualities an Integrable Systems} for  fruitful discussions  and
contributions. The author is partially supported by the scientific
research project (BAP) 2012BSP014, Pamukkale University, Denizli,
Turkey.


\vskip 1cm


\end{document}